\documentstyle[amssymb,aps,epsfig,multicol]{revtex}

\begin{document}
\title{Electron spin resonance study of Na$_{1-x}$Li$_x$V$_2$O$_5$}
\author{M.~Lohmann$^1$, H.-A.~Krug von Nidda$^1$, A.~Loidl$^1$,\\ E.~Morr\'{e}$^2$, M.~Dischner$^2$, C.~Geibel$^2$}
\address{$^1$Experimentalphysik V, Universit\"{a}t Augsburg, D-86135 Augsburg, Germany\\
 $^2$Max-Planck-Institut f\"{u}r chemische Physik fester Stoffe, D-01187 Dresden, Germany}
\date{\today}
\maketitle

\begin{abstract}
We measured X-band electron-spin resonance of single crystalline sodium vanadate
doped with lithium, Na$_{1-x}$Li$_x$V$_2$O$_5$ for $0\leq x \leq 1.3 \%$. The
phase transition into a dimerized phase that is observed at 34 K in the undoped
compound, was found to be strongly suppressed upon doping with lithium. The spin
susceptibility was analyzed to determine the transition temperature and the
energy gap with respect to the lithium content. The transition temperature
$T_{\rm SP}$ is suppressed following a square dependence of the lithium
concentration while the energy gap is found to decrease linearly. At high
temperatures ($T>T_{\rm SP}$) the susceptibility remains nearly independent of
doping.
\end{abstract}


\begin{multicols}{2}

\section{Introduction}
Since 1996, when Isobe and Ueda \cite{Iso:96} first reported the observation of
an exponential decrease of the susceptibility in NaV$_2$O$_5$ below 34\,K, this
material has been subject of intense investigation. The transition was first
considered to be a spin-Peierls transition similar to that observed in CuGeO$_3$
\cite{Hase:93}. This assumption was based on an early determination of the
structure by Carpy et al.~\cite{Carpy:75}, who proposed alternating chains of
V$^{4+}$ (spin 1/2) and nonmagnetic V$^{5+}$. This picture was able to explain
the physical properties above the transition, like the susceptibility that
closely follows that of a one-dimensional spin 1/2 Heisenberg antiferromagnet as
calculated by Bonner and Fisher \cite{Bon:64} or more recently by Eggert et al.
\cite{Egg:94}. It could not explain most of the experimental findings connected
with the transition itself nor the low-temperature state: the ratio of the
energy gap $\Delta(0)$ to the transition temperature was found to be much larger
than the expected mean-field value of $2\Delta/k_{\rm B}T_{\rm SP}=3.53$
\cite{Fuji:97}; the entropy of the jump in the specific heat is also much higher
than expected \cite{Hem:98}; and in thermal-expansion measurements two
transitions close to each other were observed \cite{Koep:98}. In the
low-temperature phase satellite reflections were reported in X-ray measurements
corresponding to a doubling of the unit cell in $a$ and $b$ and a quadrupling in
$c$ direction \cite{Fuji:97,Ravy:98}.\\ However, recent structural
investigations \cite{Smo:98,Meet:98,vonSchn:98} have shown that instead of the
originally proposed non-centrosymmetric space group $P2_1mn$, the structure of
NaV$_2$O$_5$ at room temperature has to be described by the centrosymmetric
space group $Pmmn$. In this structure only one kind of vanadium sites exists
with an average vanadium valence of V$^{+4.5}$. NaV$_2$O$_5$ can therefore be
regarded as a quarter-filled ladder system with one electron per rung. This
excludes the possibility of a simple spin-Peierls transition in this material.
The occurrence of a charge-ordering transition followed by a dimerization is
discussed \cite{Thal:98,Seo:98,Most:98}. Different types of low temperature
structures were proposed. Whereas theoretical models mainly discuss an inline or
a zig-zag ordering, a recent determination of the low-temperature structure
suggests a separation into modulated and unmodulated vanadium ladders
\cite{Lued:99}.\\ The first ESR measurements of NaV$_2$O$_5$ were carried out in
1986 by Ogawa et al.\cite{Oga:86}. Due to a large Curie contribution in the
susceptibility they did not observe the characteristic decrease below $34\,$K.
The discovery of the transition by Isobe and Ueda stimulated many other ESR
studies in this compound
 \cite{Loh:97,Vas:97,Schm:98,Yam:98}.
In this article we present electron-spin resonance (ESR) results of single
crystalline Na$_{1-x}$Li$_x$V$_2$O$_5$ for $x=0, 0.15\%, 0.3\%, 0.5\%, 0.7\%$,
$0.9\%$, and $1.3\%$ in the temperature range 4.2\,K -- 700\,K. We discuss the
ESR linewidth and the signal intensity that is directly proportional to the spin
susceptibility. Assuming a mean-field like dependence of the energy gap
$\Delta(T)$ that opens below the transition, we determine the value of the
energy gap at zero temperature and the transition temperature as a function of
the lithium concentration.

\section{Sample preparation and experiment}
The samples were small single crystals, prepared from a NaVO$_3$
flux \cite{Iso:97}. In a first step a mixture of Na$_2$CO$_3$ and
V$_2$O$_5$ is heated up to 550¬ C in air to form NaVO$_3$. In a
second step the NaVO$_3$ is mixed with VO$_2$ in the ratio of 8:1
and then heated up to 800¬ C in an evacuated quartz tube and
cooled down at a rate of 1\,K per hour. The excess NaVO$_3$ was
dissolved in water. The doped Samples were produced by
substituting in the first step Na$_2$CO$_3$ by Li$_2$CO$_3$.
However, due to a low distribution coefficient during the flux
growth process, the real amount of Li in the sample is much lower.
The real cation composition was determined in two doped samples
using inductive coupled plasma for the V content and atomic absorption spectroscopy
for the Li and Na content (see
table \ref{tab}). The result shows that the real Li content is a factor of
7.5 lower than the nominal one. For the other samples the Li
concentration was scaled accordingly, as given in table \ref{tab}. All
the samples were investigated using X-ray powder diffraction. Only
at high Li-content, a small decrease of the c lattice parameter
was observed.
\end{multicols}
\begin{table}
\begin{center}
\caption{Composition and lattice parameter of the investigated samples}
\label{tab}
\begin{tabular}{llllll}
Nominal    & resulting      &          &           &            &            \\
Li-content & Li-content     & a (\AA)  & b (\AA)   & c (\AA)    & V(\AA$^3$) \\
 (\%)      & (\%)           &          &           &            &            \\ \tableline
   0       &     0          & 11.312(3)& 3.6106(9) & 4.8031(10) &  196.17(12)\\ \tableline
   1       & 0.15 (scaled)  & 11.307(3)& 3.6095(8) & 4.8014(7)  &  195.96(9) \\ \tableline
   2.4     &  0.3 (scaled)  & 11.312(1)& 3.6112(11)& 4.8012(3)  &  196.13(7) \\ \tableline
   3.7     &  0.5 (scaled)  & 11.316(2)& 3.6123(7) & 4.8033(4)  &  196.35(7) \\ \tableline
   5       &  0.7 (measured)& 11.314(3)& 3.6103(9) & 4.8018(7)  &  196.13(11)\\ \tableline
   7       &  0.9 (scaled)  & 11.312(2)& 3.6096(11)& 4.7974(7)  &  195.88(10)\\ \tableline
  10       &  1.3 (measured)& 11.313(5)& 3.6104(19)& 4.7927(15) &  195.75(22)\\
\end{tabular}
\end{center}
\end{table}
\begin{multicols}{2}

The ESR measurements were performed using a Bruker Elexsys 500 CW spectrometer
at X-band frequency (9.48 GHz). In the temperature range 4.2 -- 300\,K a
continuous flow He-cryostat (Oxford Instruments) and between 300\,K and 700\,K a
nitrogen cryostat (Bruker) was used. The samples were orientated in a way that
the applied external field was always perpendicular to the crystallographic $b$
axis and could be rotated about this axis. All measurements were made at the
orientation with the narrowest resonance line, i.e.~the external field $H$ being
parallel to the $a$ axis.

\section{Electron-spin resonance}
\begin{figure}[htb]
\begin{center}
\centerline{\epsfig{file=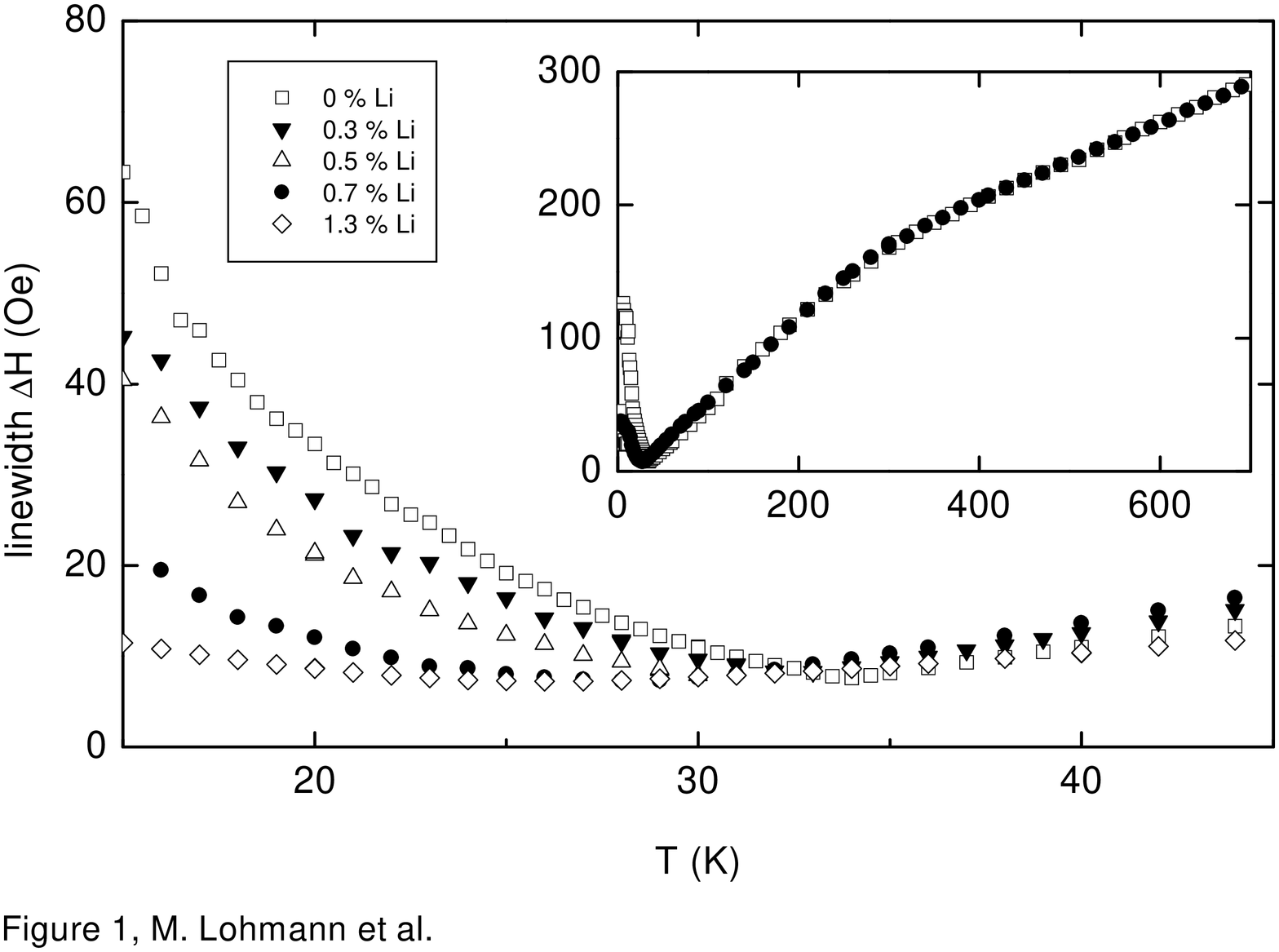,angle=-90,clip,width=8cm}} \vspace{10pt}
\caption[ESR linewidth]{ESR linewidth of Na$_{1-x}$Li$_x$V$_2$O$_5$ below 50\,K for different lithium concentrations,
the inset shows the linewidth for x=0 (open squares) and 0.7 \% lithium (filled circles).}
\label{dh}
\end{center}
\end{figure}

NaV$_2$O$_5$ shows one single lorentzian-shaped resonance line with an anisotropic $g$ value between
1.976 ($H$ parallel $a$ axis) to 1.936 ($H$ parallel $c$ axis) \cite{Loh:97}. At high temperatures the linewidth of this resonance
decreases monotonically with decreasing temperature and is independent from lithium doping as shown in the inset of figure
\ref{dh} for the undoped and the 0.7 \% lithium doped sample. Below 34 K the linewidth increases
again. This increase was found to be rather strongly suppressed by doping (figure \ref{dh}). While the linewidth in the undoped sample
increases by a factor of 4 from 34\,K down to 15\,K, for 1.3 \% lithium content the increase is only about 40
\%. This clearly indicates that the increase of the linewidth below 34\,K is directly connected to the
transition, which is suppressed upon lithium doping as will be shown below.
In the whole temperature range the ESR signal is strongly exchange narrowed and
no hyperfine structure due to the $^{51}$V-spin ($I=7/2$) is observed
\cite{Sper:74}.

\begin{figure}[tb]
\begin{center}
\centerline{\epsfig{file=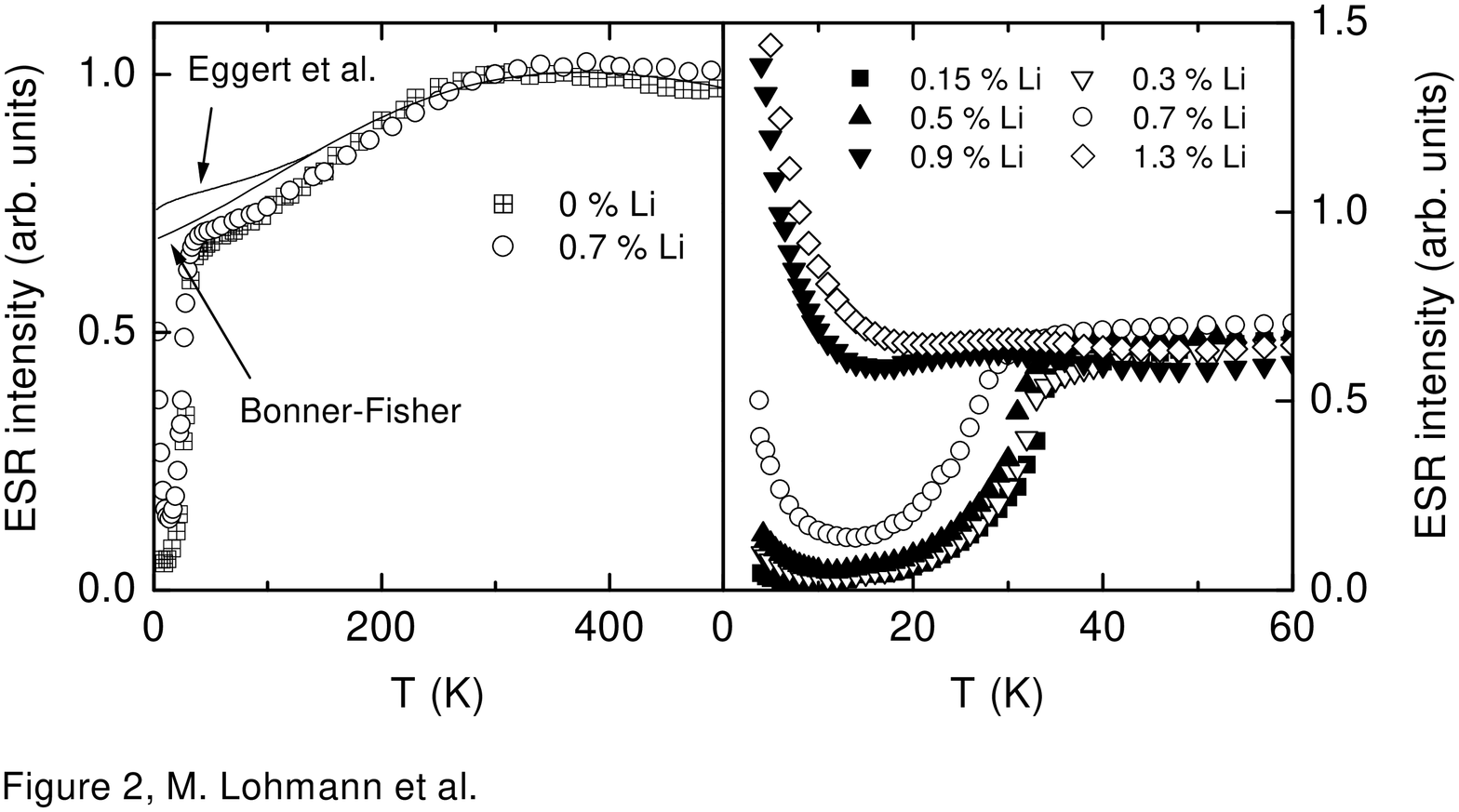,angle=-90,clip,width=8cm}} \vspace{10pt}
\caption[ESR linewidth]{Spin susceptibility of Na$_{1-x}$Li$_x$V$_2$O$_5$:\\
a) (left) x=0 (squares) and 0.7 \% lithium (circles),
the solid lines represent the fits using the theory of Bonner and Fisher \cite{Bon:64}
and Eggert et al. \cite{Egg:94} with $J=578$\,K.\\
b) (right) susceptibility below 60\,K for different lithium concentrations.}
\label{int_li}
\end{center}
\end{figure}

We therefore propose that the broadening of the linewidth
below the transition appears because the exchange narrowing becomes less effective, probably due charge localisation.\\
A similar overall temperature dependence of the linewidth is observed in CuGeO$_3$ \cite{Yam:96}.
Yamada et al.~qualitatively explained the high-temperature behavior in both CuGeO$_3$ and NaV$_2$O$_5$ by identifying
the anisotropic Dzyaloshinsky-Moriya exchange interaction $H_{\rm DM}$ as the dominating interaction responsible for the
linebroadening \cite{Yam:98,Yam:96}.
The Dzya\-lo\-shin\-sky-Moriya interaction is given by $\sum_i d_{ii+1}\cdot (\mathbf{S}_i \times
\mathbf{S}_{i+1})$ for neighboring spins $\mathbf{S}$, were $d_{ii+1}$ can be estimated as $d_{ii+1}\simeq(\Delta
g/g)|J|$ \cite{Moriya:60}. We found that both $g$ value and exchange coupling
constant $J$ (that can be determined from the spin susceptibility, see fig. \ref{int_li}a)
remain nearly unaffected by doping. This is consistent with the fact that no
concentration dependence of the linewidth was detected at high temperatures.

We also determined the spin susceptibility of Na$_{1-x}$Li$_x$V$_2$O$_5$ from
the intensity of the ESR signal. Since it is difficult to determine the absolute
values of the susceptibility by ESR, only relative values are given, the curves
being scaled to one at 300\,K. An estimation of the absolute intensity is
consistent with one vanadium per formular unit contributing to the signal. As
mentioned before, the spin susceptibility above the transition is nearly
insensitive to lithium doping. In figure \ref{int_li}a the undoped sample is
compared with the 0.7 \% lithium doped sample. For $T>200$\,K both curves nicely
agree with the theoretical fit using the dependence calculated by Bonner and
Fisher \cite{Bon:64} or Eggert et al.~\cite{Egg:94} with $J=578$\,K. Both
calculations give the same results above $T=0.3 J\simeq 175$\,K. Below this
temperature the more exact calculation of Eggert et al.~shows an even more
pronounced disagreement with the data. The reason for this deviation is not
totally resolved. It could be due to a dimensional crossover as was suggested
from X-ray investigations (Ravy et al. \cite{Ravy:98} predict a deviation from
the Bonner-Fisher theory up to temperatures much higher than 90\,K) or due to
the existence of structural fluctuations.

Figure \ref{int_li}b displays the spin susceptibility below 60\,K for different lithium
concentrations. The transition shifts to lower temperatures and the decrease
of the susceptibility becomes less pronounced with increasing lithium content.
We also observe a Curie like increase at lowest temperatures that
increases with doping. In the sample Na$_{1-x}$Li$_x$V$_2$O$_5$ with $x=1.3\%$
the transition is no longer visible (see figure \ref{int_li}b).\\

\begin{figure}[htb]
\begin{center}
\centerline{\epsfig{file=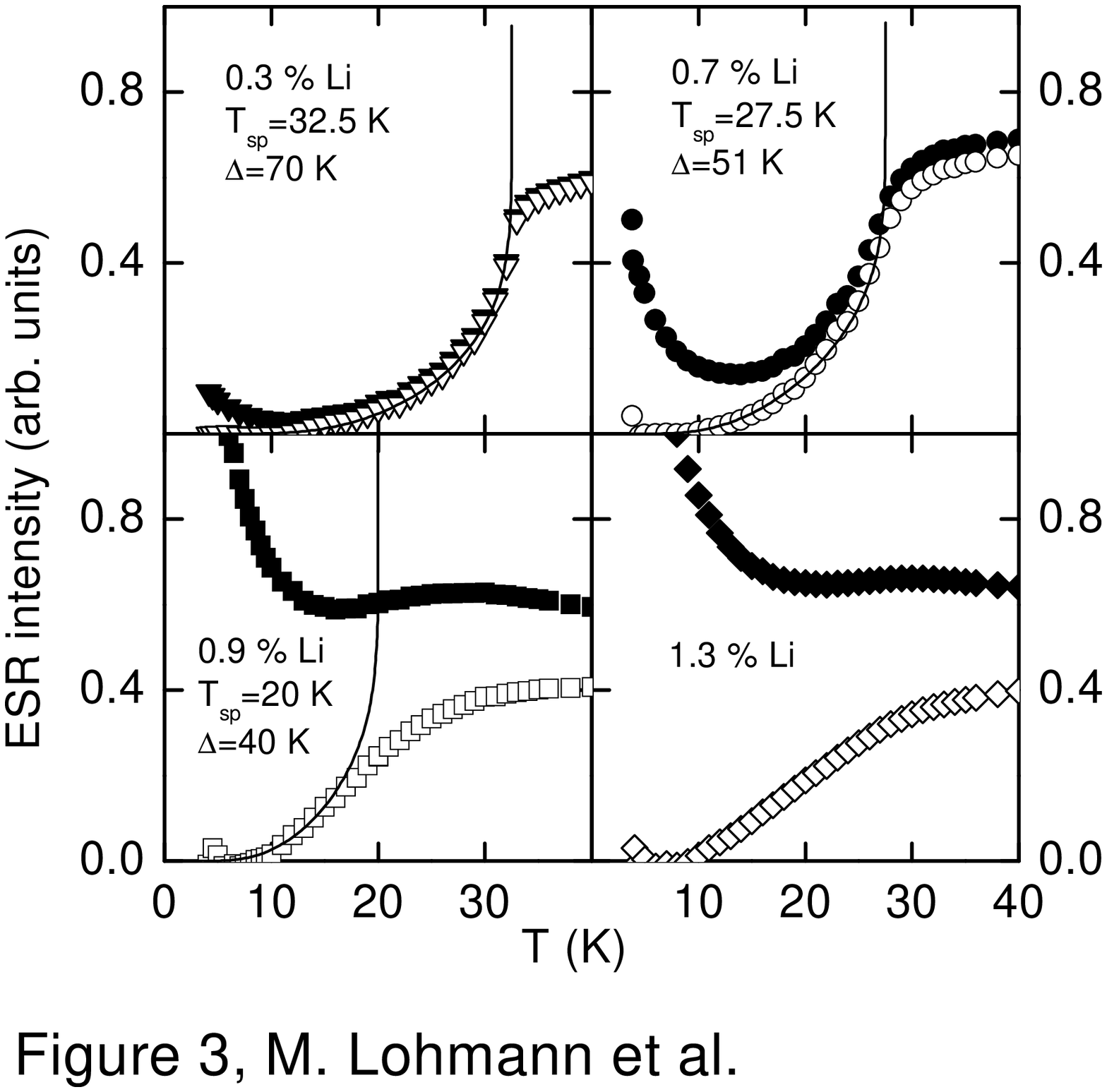,angle=0,clip,width=8cm}} \vspace{10pt}
\caption{Spin susceptibility of Na$_{1-x}$Li$_x$V$_2$O$_5$ for different lithium concentrations.
The filled symbols are original data, the open symbols
represent the data after substraction of the Curie contribution, the solid
lines show the fits assuming a mean-field like energy gap $\Delta(T)$.}
\label{fit}
\end{center}
\end{figure}

To analyse the data, a Curie law was fitted to the data points below 10\,K and subtracted. The curves were then
analyzed using a mean-field like temperature dependence of the energy gap and $\chi(T)\propto
\rm{exp}(2\Delta/k_{\rm B}T)$. For the temperature dependence of the energy gap $\Delta(T)$ the exact mean-field values were taken; $\Delta(0)$, and
$T_{\rm SP}$ being the only fitting parameters. In this case it is preferable to use this method rather than
fitting with the theory of Bulaevskii \cite{Bul:69} because the uncertainty at low temperatures caused by the Curie
contribution strongly influences the determination of the energy gap $\Delta(0)$. Examples of the fitting procedure
for different $x$ are given in figure \ref{fit}. In the samples with $x\leq 5\%$ perfect agreement of the data and the
fitting curves is found. The transition is broadened with increasing lithium content thus causing an increasing
uncertainty for the high doped samples $x=0.9\%$ and $x=1.3\%$. While a determination of $\Delta(0)$, and $T_{\rm SP}$
is still possible in the $x=0.9\%$ lithium doped sample, in the 1.3\% doped sample no clear choice of $\Delta(0)$ and
$T_{\rm SP}$ could be made, because the phase transition is strongly broadened in temperature and it is not clear
how to determine the Curie contribution exactly (if the data are treated like those of the other samples assuming
that a low temperature only the Curie contribution exists this contribution is probably overestimated leading to a
seemingly linear decrease of the susceptibility as shown in figure \ref{fit}). \\

\begin{figure}[htb]
\begin{center}
\centerline{\epsfig{file=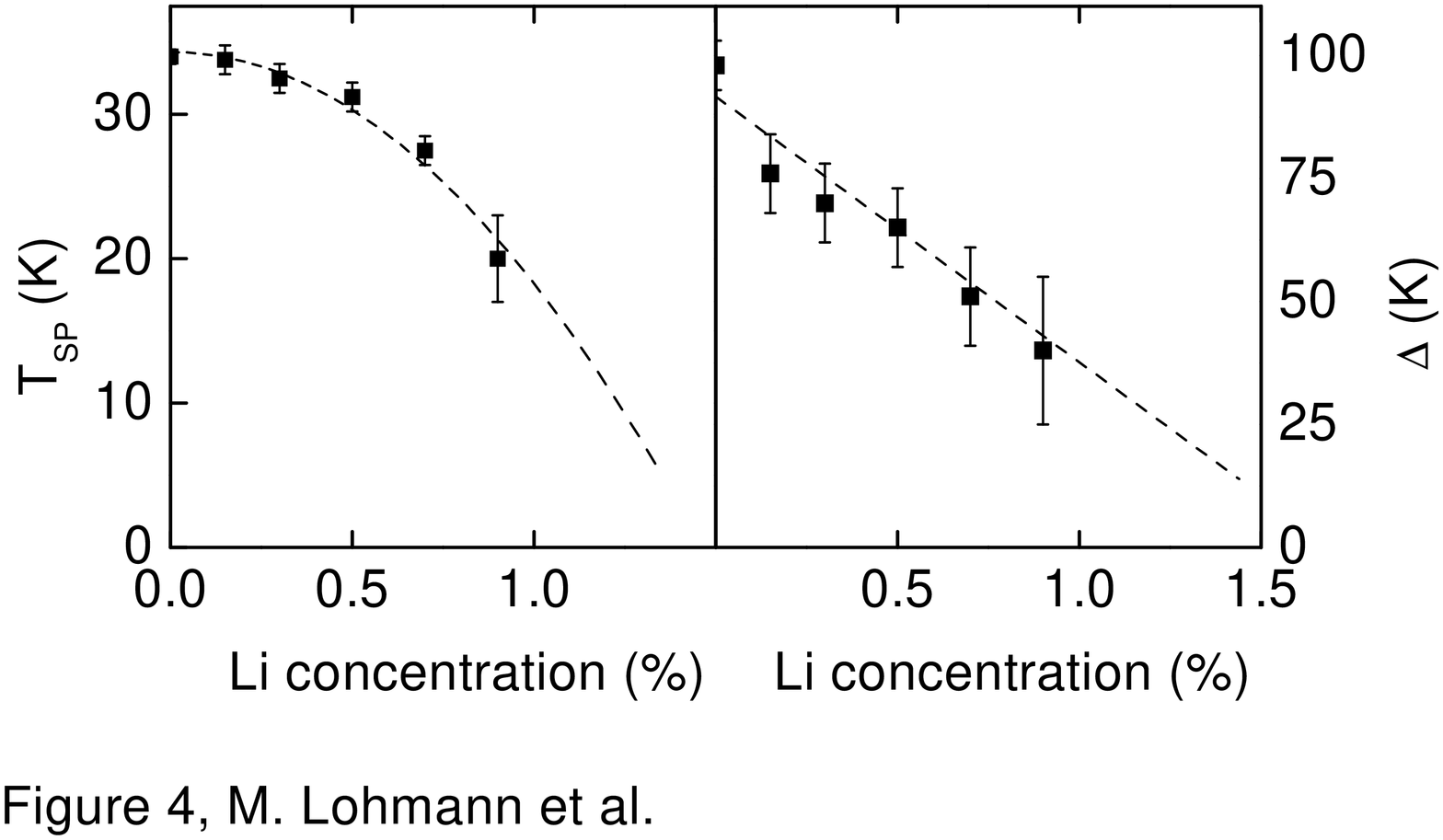,angle=-90,clip,width=8cm}} \vspace{10pt}
\caption{Variation of the transition temperature $T_{\rm SP}$ and energy gap $\Delta(0)$ with lithium concentration.
The dashed line in the left graph represents a fit according to $T_{\rm SP}(x)\propto a-bx^2$, the dashed line in the
right graph is a guide to the eye.}
\end{center}
\label{T_d}
\end{figure}

The results for the transition temperature $T_{\rm SP}$ and the
energy gap $\Delta(0)$ are displayed in figure \ref{T_d}. The transition
temperature is seems to follow a $T_{\rm SP}\propto a-bx^2$
function (dashed line). The energy gap $\Delta(0)$ varies linearly
with the lithium content. However, since the errors in the
determination of the lithium content have to be taken into
account, further investigation is necessary to confirm the exact
dependencies. For both cases the value of the assumed functions
differs from zero (i.e. no transition occurs) at $x=1.3 \%$
lithium. This suggests that even in the case of 1.3 \% lithium
doping the transition is not completely suppressed. Another
interesting result is that the ratio $2\Delta/k_{\rm B}T_{\rm SP}$
decreases from the strong coupling value of 5--6 in undoped
NaV$_2$O$_5$ to values close to the mean field result of 3.53, i.e.~3.7--4 in the samples with $x=0.5\%$ and
$x=0.7\%$.\\

\section{Conclusions}
In conclusion we have presented ESR results on Na$_{1-x}$Li$_x$V$_2$O$_5$ for $0
\leq x \leq 1.3 \%$. The linewidth and the spin susceptibility above the
transition were found to be nearly independent from the lithium concentration.
At low temperatures the increase of the linewidth is suppressed with growing
lithium content. The spin susceptibility was analyzed using a mean-field like
model to extract the transition temperatures and the $T=0$ value of the energy
gap with respect to the doping. It was found that the transition temperature and
the energy gap decrease monotonically on increasing Li concentration, suggesting
a square dependence of the transition temperature and a linear decrease of the
energy gap. Considering these dependencies it is highly probable that even in
the highest doped sample a transition still persists.\\ Although there is no
theoretical prediction for the suppression of the transition upon doping in
NaV$_2$O$_5$, one can speculate about the relevant physical properties. The
lattice parameters (table 1) show only a slight doping dependence. It is
consequently very improbable that the suppression of the transition can be
explained with the change of the lattice. In a normal spin-Peierls system the
transition depends on the spin-phonon coupling $g$ and the phonon frequency
$\omega$ \cite{Uhr:98}. The transition temperature should be in the order of
$g/\omega^2$. Substitution of the lighter lithium ions for sodium is expected to
increase the phonon frequency $\omega$ thus reducing the transition temperature.
This scenario could explain the monotonic decrease of the transition temperature
upon doping. In this context a direct observation of the phonon frequencies in
lithium-doped samples would be very interesting.

In Na$_{1-x}$Li$_x$V$_2$O$_5$ the lithium ions are located on the off-chain
sodium positions. In contrast to CuGeO$_3$ doped off-chain with silicon
\cite{Ren:95}, where antiferromagnetic order appears for concentrations as low
as 0.5\%, no signs of magnetic order were found . In CuGe$_{1-x}$Si$_x$O$_3$ the
spin-Peierls transition decreases linearly as $T_{\rm SP}(x)\propto a-bx$
\cite{Ren:95}. While in CuGeO$_3$ off-chain substitutions (like Si
\cite{Ren:95}) and in-chain substitutions (like Zn \cite{Lem:97,Oser:95} or Mg
\cite{Hase:95}) have been extensively studied, in NaV$_2$O$_5$ many interesting
work in this field remains to be done.

We gratefully acknowledge helpful discussion with A.~Kampf. This work was partly
supported by BMFT under contract no. 13N6917/0 and DFG
 under contract no. 20 264/10-1.\\

\end{multicols}

\end{document}